\documentclass[aps, prl, twocolumn,showpacs,preprintnumbers,amsmath,amssymb]{revtex4}

\usepackage{graphicx}
\usepackage{dcolumn}
\usepackage{bm}

\begin{document}

\preprint{}

\title{How much security does Y-00 protocol provide us?}

\author{Tsuyoshi NISHIOKA}
 \email{nishioka@isl.melco.co.jp}
 \affiliation
 {Information Technology R \& D Center, Mitsubishi Electric 
Corporation\\
5-1-1 Ofuna, Kamakura, Kanagawa 247-8501, JAPAN\\
TEL: +81-467-41-2190 \quad FAX: +81-467-41-2185}
\author{Toshio HASEGAWA}%
 \email{toshio@isl.melco.co.jp}
 \affiliation
 {Information Technology R \& D Center, Mitsubishi Electric 
Corporation\\
5-1-1 Ofuna, Kamakura, Kanagawa 247-8501, JAPAN\\
TEL: +81-467-41-2190 \quad FAX: +81-467-41-2185}

\author{Hirokazu ISHIZUKA}
 \email{ishizuka@isl.melco.co.jp}
 \affiliation
 {Information Technology R \& D Center, Mitsubishi Electric 
Corporation\\
5-1-1 Ofuna, Kamakura, Kanagawa 247-8501, JAPAN\\
TEL: +81-467-41-2190 \quad FAX: +81-467-41-2185}

\author{Kentaro IMAFUKU}
 \email{imafuku@imailab.iis.u-tokyo.ac.jp}
 \affiliation
 {Institute of Industrial Science, University of Tokyo \\
4-6-1 Komaba, Meguro-ku, Tokyo 153-8505, JAPAN\\
TEL: +81-3-5452-6232 \quad FAX: +81-3-5452-6631}

\author{Hideki IMAI}
 \email{imai@iis.u-tokyo.ac.jp}
 \affiliation
 {Institute of Industrial Science, University of Tokyo \\
4-6-1 Komaba, Meguro-ku, Tokyo 153-8505, JAPAN\\
TEL: +81-3-5452-6232 \quad FAX: +81-3-5452-6631}


\begin{abstract}
New quantum cryptography, often called Y-00 protocol, 
has much higher performance than the conventional quantum 
cryptographies.
It seems that the conventional quantum cryptographic attacks
are inefficient at Y-00 protocol
as its security is based on the different grounds
from that of the conventional ones.
We have, then, tried to cryptoanalyze Y-00 protocol
in the view of cryptographic communication system.
As a result, it turns out that the security of Y-00 protocol is
equivalent to that of  classical stream cipher. 
\end{abstract}

\maketitle

%
%
Quantum cryptography has appeared as a promising way to achieve security
without depending on any computational complexity assumption. However, most of
the proposed schemes up to date are based on single photon states (QCSPS)
thus, presenting a well-known negative characteristic, namely, its bit rate is
much slower than that of normal optical communication systems.

Recently a quantum cryptography scheme which uses mesoscopic coherent states
has been reported \cite{BCKY03,BCKYDPP02,Ba02,HKS02}.
This scheme has much higher
performance than conventional ones which are based on single photon
states\cite{BB84,Be92,E91}. Because this scheme based on mesoscopic coherent
states, often called ``Y-00 protocol\cite{HKSU03},'' has an average photon
number of 100-1,000 photons per pulse, its bit rate is expected to be
100-1,000 times faster than that of QCSPS. In addition, the required
technical level to realize the protocol is supposed to be quite the same as in
conventional optical systems. Y-00 protocol would be, then, a sufficiently
fast and easily realizable quantum cryptography scheme, if it had actually
perfect security.

\begin{figure}[ptbh]
\includegraphics
[viewport = 50  380 550 600, clip=true, width=1.0\linewidth]
{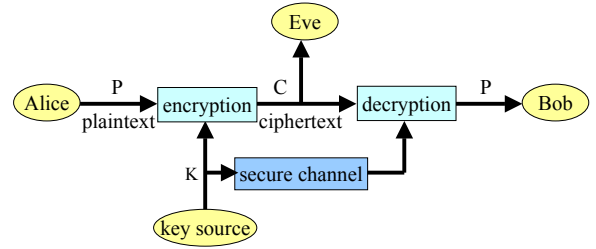}\caption{\textit{Cryptographic communication system}}%
\label{fig:cryptocom}%
\end{figure}

In this paper, we show that the Y-00 protocol does not provide perfect
security, even against the simplest of cryptographic attacks,
ciphertext-only ones. A usual cryptographic system consists of two
channels (Fig.\ref{fig:cryptocom}):
an open channel for exchanging encrypted messages
and a secure channel for key distribution.
%
%
%
%
Quantum cryptography based protocols, including Y-00 protocol, 
provide a realization of
the secure channel for key distribution. Let us underline that our attack
targets are not only the secure channel, but also the open channel for messages.

We also show that the security of Y-00 protocol is just
equivalent to that of a classical stream cipher. In other words, we can safely
say that Y-00 protocol has no perfect security and that its
security depends on just a common computational complexity assumption, being
then no better than currently used schemes.

%
%

Y-00 protocol is a quantum key expansion (QKE) scheme. Both Alice and Bob must
share a secret key $K_{s}$ in advance. Notice that, even conventional quantum
cryptography can be regarded as quantum key expansion schemes, since a short
key is needed for the authentication of the classical channel. Y-00 protocol
uses the secret key for quantum modulation and de-modulation in a quantum
channel for expansion of the key.

Y-00 protocol has, then, $2M$ non-orthogonal coherent states called ``qumodes''
:
\begin{equation}
|\alpha e^{i\theta_{k}}\rangle,\quad\theta_{k}=\pi k/M,
\end{equation}
where $\alpha\in{\bm C}$ and $k\in\{0,\ldots,2M-1\}$, or
\begin{equation}
|\Psi(\theta_{k})\rangle= |\alpha\cos\theta_{k}/2\rangle_{H}|\alpha\sin
\theta_{k}/2\rangle_{V}, \label{qumode}%
\end{equation}
in the polarization coding where the suffix ``H'' and ``V'' are horizontal and
vertical polarization modes. We focus on the polarization coding hereafter.

The $2M$-states are divided into $M$-pairs:
\begin{equation}
|\alpha\cos\theta_{k}/2\rangle_{H}|\alpha\sin\theta_{k}/2\rangle_{V},
\end{equation}
and
\begin{equation}
|\alpha\cos\theta_{M+k}/2\rangle_{H}|\alpha\sin\theta_{M+k}/2\rangle_{V} ,
\end{equation}
where $k\in\{0,\ldots,M-1\}$ and the pairs are orthogonal to each other. A
specific pair, then, determines a specific polarization base.

At first Alice and Bob generate a pseudo-random number stream $k_{i}%
\in\{0,\ldots,M-1\}$ from the secret key $K_{s}$ in synchronized manner:
\begin{equation}
PRNG:K_{s}\rightarrow k_{i}, \quad i\in{\bm N},
\end{equation}
where $PRNG$ is a pseudo-random number generator and the number $k_{i}$
determines the polarization base. Alice generates a number $r_{i}\in\{0,1\}$
with a physical random number generator (PhRNG) and modulates the $r_{i}$ into
a qumode in the $k_{i}$-base. Bob observes the qumode sent from Alice with the
$k_{i}$-base and de-modulates $r_{i}$ from the qumode. The mechanism is called
``M-ary level ciphering.''

Y-00 protocol has another important mechanism called ``Ciphering Wheel.''
Ciphering Wheel is a rule for assigning bits 0 and 1 of $r_{i}$ to two qumodes
on each base. Two closest qumodes in the neighboring bases are generally
assigned opposite bit values. Thus one must distinguish a correct qumode to
get a correct bit. However, because of the fundamental quantum fluctuation in
any measurement by eavesdroppers, the discrimination of the qumode is
impossible for eavesdroppers when $M$ is sufficiently large, unless they know
the information of $k_{i}$.

\begin{figure}[ptbh]
\includegraphics
[viewport = 50  230 550 640, clip=true, width=1.0\linewidth]
{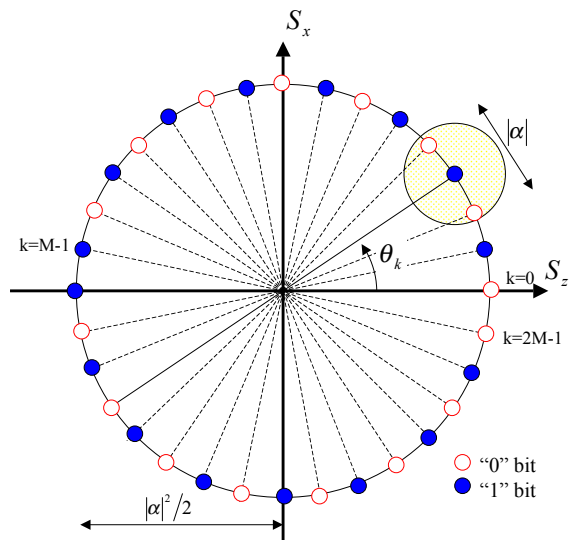}
\caption{\textit{qumode in Poincar\'{e} sphere at $M=16$}}%
\label{fig:qumode}%
\end{figure}

These circumstances can be easily understood with the so-called Poincar\'{e}
representation. The qumode is represented by a point on the Poincar\'{e}
sphere and all qumodes are located on the equator including the $z$-axis and
$x$-axis defined by the Stokes operators in Fig.\ref{fig:qumode}:
\begin{equation}
S_{z}=\frac{1}{2}(a^{\dagger}a-b^{\dagger}b),\quad\mbox{and}\quad S_{x}%
=\frac{1}{2}(a^{\dagger}b+b^{\dagger}a),
\end{equation}
where $a$ is an annihilation operator for the horizontal mode and $b$ is an
annihilation operator for the vertical mode.
%
%
%
%
The qumode (\ref{qumode}) is, then, represented by the point
\begin{equation}
(S_{z},S_{x})=\frac{1}{2}|\alpha|^{2}(\cos\theta_{k},\sin\theta_{k}),
\end{equation}
and it has the following isotropic quantum fluctuation
\begin{equation}
\Delta S_{z}=\Delta S_{x}=\frac{1}{2}|\alpha|.
\end{equation}
$N_{\sigma}$-other qumodes are included in the fluctuation where $N_{\sigma
}=M/(\pi|\alpha|)$ and no one can distinguish the correct qumode from the
others. Therefore nobody except Alice and Bob draw out the correct bit
\textit{whereas Bob's decision has to be made only between two nearly
orthogonal state in the same basis defined by a given $k_{i}$}~\cite{BCKY03}.

From the above argument, one can see that Alice and Bob can surely safely
share the new random bits $\{r_{i}\}$ whose length is longer than the original
shared random common key $K_{s}$. Alice and Bob seem to be able to
realize QKE by using the Y-00 protocol. However, a careful consideration tells
us that this protocol doesn't help Alice and Bob to expand their key in a
perfect secure way.

Let us imagine that Eve classifies the bases into two classes: The one class
$C_{+}$ consists of bases which has the same bit-assignment as the $k_{i}=0$
base and the other class $C_{-}$ consists of the rest of the bases. She can,
then, define a mono-bit mapping $CW(\cdot)$ from the base $k_{i}$:
\begin{equation}
CW:k_{i}\rightarrow\tilde{k}_{i}=\left\{
\begin{array}
[c]{@{\,}ll}%
0 & \mbox{if $k_{i}\in C_{+}$}\\
1 & \mbox{if $k_{i}\in C_{-}$}%
\end{array}
\right.  .
\end{equation}

Since Eve seizes the qumode and its base vaguely, she, then, selects an
appropriate base which belongs to the class $C_{+}$ from the candidates and
gets a bit $l_{i}$ from observation under the selected base. Since the
candidates are not always the same as the true bases, the bit $l_{i}$ cannot
always be equal to $r_{i}$. However one will find that the following important
relation holds:
\begin{equation}
l_{i}=r_{i}\oplus\tilde{k}_{i}. \label{keystream}%
\end{equation}
Notice that the quantum error due to the miss choice of base is absorbed in
the second term of the right side. Moreover, since the inner product of the
true qumode $|\Psi(\theta_{k})\rangle$ and the orthogonal qumode on the
incorrect base $k+\Delta k$ is given by
\begin{equation}
|\langle\Psi(\theta_{k})|\Psi(\theta_{M+k+\Delta k})\rangle|^{2}%
=e^{-2|\alpha|^{2}(1+\sin(\pi\Delta k/2M))},%
\end{equation}
(the product is extremely small if the qumode is a mesoscopic state where the
order of $\Delta k$ is at most $N_{\sigma}$), the error derived from M-ary
ciphering mechanism is negligible.

In addition, Eve can make her measurement undetectable by Alice and Bob,
by resending a similar quantum state to Bob who is interested in only the
discrimination of the possible two states. In principle, Bob can check if the
state is really one of two possible state, if the quantum channel can be
assumed free from noise. However, it is much harder to check it than only to
distinguish from each other. Moreover, it requires much higher technology than
what is assumed in the original Y-00 protocol. Also, the assumption that the
quantum channel is free from noise is a quite unrealistic one.

Again, we would like to stress that the detection of Eve's measurement is
impossible for Alice and Bob under the relevant assumptions assumed in the
Y-00 protocol, paying serious attention to the philosophy of the original
proposal, that is, an easily realizable protocol with conventional optical
technologies. Also, note that signal amplification, which is considered as one
of the mains advantages of the protocol, is impossible under the assumption
that Eve's activities can be detected.

\begin{figure}[ptbh]
\includegraphics
[viewport = 50  235 550 570, clip=true, width=1.0\linewidth]
{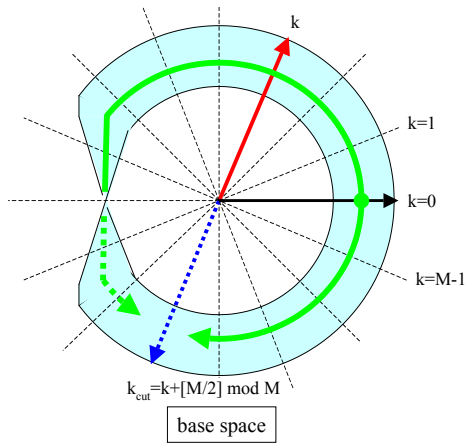}\caption{\textit{Local Classification on bases at $M=16$}}%
\label{fig:localclass}%
\end{figure}

There is a small technical problem that the classification is not globally
well-defined because the base space is topologically homomorphic to the
M\"{o}bius ring. The vertical polarization mode on the $k=0$ base is adjacent
to the horizontal one on the $k=M-1$ base, though the base space has cyclic
structure of a module $M$. However one can solve this problem, by introducing the
following ``local'' classification in the neighborhood of a given $k$-base in
Fig.\ref{fig:localclass}:

\begin{enumerate}
\item  Fix the most far base $k_{\mathrm{cut}}\equiv k+[M/2]\bmod M$ as the
cut base if a certain $k$ is given.

\item  Classify each bases from $k=0$ to $k_{\mathrm{cut}}$ in both additive
and subtractive directions.
\end{enumerate}

The local classification is well-defined in the neighborhood of $k$ and unique
on any $k$. It is no problem that the neighborhood of $k_{\mathrm{cut}}$ is
ill-defined because the qumodes on the $k$-base and $k_{\mathrm{cut}}$ are
orthogonal to each other, i.e.
\begin{equation}
|\langle\Psi(\theta_{k})|\Psi(\theta_{k_{\mathrm{cut}}}\rangle|^{2} \simeq
e^{-(2-\sqrt{2})|\alpha|^{2}}\rightarrow0. \label{cutstate}%
\end{equation}
%
%
%
%

Let us recall the important fact that, in the Y-00 protocol, Eve can get
(\ref{keystream}) without being detected by Alice and Bob. First of all, one
should notice that this situation can be simulated in a classical way.

\begin{figure}[ptbh]
\includegraphics
[viewport = 50  400 550 650, clip=true, width=1.0\linewidth]
{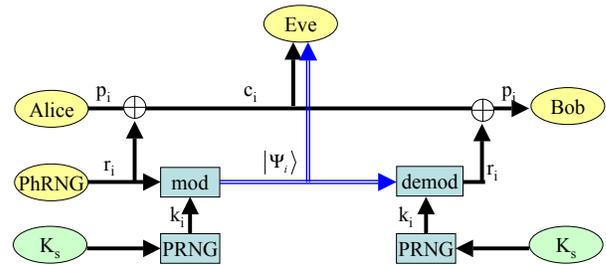}\caption{\textit{Y-00 communication system}}%
\label{fig:Yuencom}%
\end{figure}

Let us investigate the Y-00 protocol as a cryptographic communication system
where one-time pad is used as encryption algorithm for messages in
Fig.\ref{fig:Yuencom}.
%
%
%
%

Eve gets two clues to cryptanalysis: One clue is a ciphertext $c_{i}$ of the
message:
\begin{equation}
c_{i}=p_{i}\oplus r_{i}, \label{cipher}%
\end{equation}
where $p_{i}$ is a plaintext of the message which is legible to anybody and
has often language characteristics. The other clue is an inaccurate qumode
that Eve observes in the quantum channel.

As Eve knows the two bit streams (\ref{cipher}) and (\ref{keystream}) in
this way, she gets another bit stream
\begin{equation}
c_{i}\oplus l_{i}=p_{i}\oplus\tilde{k}_{i}. \label{streamcipher}%
\end{equation}
The bit stream is nothing but a classical stream cipher where $c_{i}\oplus
l_{i}$ is a ciphertext, $\tilde{k}_{i}$ is a key stream, and its generator
algorithm is $CW\circ PRNG(\cdot)$. Therefore Y-00 protocol has no perfect
security and its security is based on computational complexity.

The above discussion is independent of the encryption algorithm. In the case
of a block cipher algorithm instead of one-time pad, we may observe
(\ref{keystream}) in a block whose size is $N$
\begin{equation}
L_{J}=R_{J}\oplus\tilde{K}_{J},
\end{equation}
where $\oplus$ is bitwise XOR and $L_{J}$, $R_{J}$, and $\tilde{K}_{J}$ are
concatenations of $l_{i}$s, $r_{i}$s, and $\tilde{k}_{i}$s, for example,
\begin{equation}
L_{J}=l_{(J-1)N+1}||l_{(J-1)N+2}||\cdots||l_{JN}.
\end{equation}
The ciphertext is given by
\begin{equation}
C_{J}=E_{R_{J}}(P_{J}),\quad\mbox{and}\quad P_{J}=D_{R_{J}}(C_{J}),
\label{blockcipher}%
\end{equation}
where $P_{J}$ and $C_{J}$ are block sequences instead of each bit streams and
$E(\cdot)$, $D(\cdot)$ are encryption and decryption algorithm. We, then, get
the following relation
\begin{equation}
P_{J}=D_{L_{J}\oplus\tilde{K}_{J}}(C_{J}), \label{bruteforce}%
\end{equation}
instead of the simple relation (\ref{streamcipher}). Brute-force attack on the
freedom of $K_{s}$ is applicable to (\ref{bruteforce}) in the worst case that
we have no clue to cryptoanalyze the encryption algorithm. Therefore its
security is never beyond that of a scheme based on computational complexity.

This attack is not applicable to QCSPS like BB84 because it has the
feature of eavesdropping detection. QCSPS with non-separable carrier abandons
the communication itself as soon as eavesdropping is detected.

It is interesting to analyze our results in comparison to the ones related to
secret key agreement over classical noisy channels. It is known that certain
noisy correlated data can provide finite secrecy capacity and perfect secure
key agreement between Alice and Bob even when \ the correlation between Alice
and Eve is less noisy than the correlation between Alice and Bob (if an
authenticated noiseless public channel is provided) \cite{Wy75,CK78,Ma93,AC93}. Y-00 protocol would be, then, expected as a practical and
efficient implementation of this noisy correlation using quantum noise. The
result (\ref{keystream}), however, tells us that the correlation implemented
by Y-00 protocol is not a stochastic one but one which comes from 
a deterministic
bit-flip channel owing to the local classification and the appropriate
base-selection. Therefore, Y-00 protocol does not provide ``genuine'' noisy
correlations between Alice, Bob and Eve and, hence, does not satisfy the
conditions specified in\cite{Wy75,CK78,Ma93,AC93}.

We assume that qumodes are mesoscopic, $|\alpha|^{2}\gg1$ in the above
discussion. If $|\alpha|^{2}$ is sufficient small, quantum effects are
expected to be no longer negligible. In our attack, if the two qumodes on the
most far base $k_{\mathrm{cut}}$ are not efficiently orthogonal to the true
qumode: $|\alpha|^{2}<1+1/\sqrt{2}$ from (\ref{cutstate}) , the clue
(\ref{keystream}) is not correct and the eavesdropping channel becomes a
stochastic noisy channel. Y-00 protocol would, then, recover perfect security
if it used microscopic coherent states instead of mesoscopic ones, but its
extremely high performance would be lost.

%

%
%
\begin{acknowledgments}
The authors thank Anderson C. A. Nascimento and Dr. Toyohiro Tsurumaru
for enlightening comments and discussions.
This work was supported by the project on ``Research and Development
on Quantum Cryptography'' of Telecommunications Advancement
Organization as part of the programme ``Research and Development on
Quantum Communication Technology'' of the Ministry of Public
Management, Home Affairs, Posts and Telecommunications of Japan.
\end{acknowledgments}

%
%
\end{document}